\documentclass[useAMS,usegraphicx]{mn2e}
\usepackage{epsfig}
\usepackage{times} 

\title[]{A powerful new method for probing the atmospheres of transiting exoplanets}

\author[I.A.G. Snellen et al.]{I.A.G. Snellen \thanks{E-mail:
ignas@roe.ac.uk}\\
Institute for Astronomy, University of Edinburgh, Blackford Hill
, Edinburgh EH9 3HJ, UK}

\begin{document}
\date{Accepted .... Received ...; in original form ....}

\maketitle
\begin{abstract}
Although atmospheric transmission spectroscopy of HD209458b with the Hubble 
Space Telescope has been very successful, 
attempts to detect its atmospheric absorption features 
using ground-based telescopes have so far been fruitless.
Here we present a new method for probing the atmospheres of transiting 
exoplanets which may be more suitable for ground-based observations, 
making use of the Rossiter effect.
During a transit, an exoplanet sequentially blocks off light from 
the approaching and receding parts of the rotating star, 
causing an artificial radial velocity wobble. The amplitude of this signal is 
directly proportional to the effective size of the transiting object, and
the wavelength dependence of this effect can reveal atmospheric absorption 
features, in a similar way as with transmission spectroscopy.
The advantage of this method over conventional atmospheric transmission 
spectroscopy is that it does not rely on accurate photometric 
comparisons of observations on and off transit, but instead depends on the 
relative velocity shifts of individual stellar absorption lines 
within the same on-transit spectra. 
We used an archival VLT/UVES data set to apply this method to  
HD209458. The amplitude of the Rossiter effect is shown to be
$1.7^{+1.1}_{-1.2}$ m/sec higher in the Sodium D lines than 
in the weighted average of all other absorption lines in the observed 
wavelength range, corresponding to an increment of 4.3$\pm$3\% (1.4$\sigma$).
The uncertainty in this measurement 
compares to a photometric accuracy of 5$\times 10^{-4}$ for conventional 
atmospheric transmission spectroscopy, more than an order of magnitude 
higher than previous attempts using ground-based telescopes.
Observations specifically designed for this method could increase the accuracy
further by a factor 2$-$3.
\end{abstract}
\section{Introduction}

\begin{table*}
\caption{ \label{log} The log of the observations, with in column 1 the 
date, in column 2 the total the exposure time, in 
column 3 the typical signal to noise ratio per exposure, in 
columns 4 and 5 the range in airmass and seeing, and 
in column 6 the orbital phase of HD209458b.}
\begin{tabular}{ccccccrl}
Data&Observing      & Exposure        & S/N & Airmass & Seeing  & Phase   & Comments \\
Set&Date           & Times (sec)     &     &         & ($''$)  &         &          \\
1&05/08/2002& 15$\times$400   & 400 & 1.407$-$1.558&  $\sim$2& $-$0.007,$+$0.021&on transit\\  
2&11/08/2002& 15$\times$400   & 535 & 2.151$-$1.394&0.9$-$1.5& $+$0.663,$+$0.693&off transit\\
3&12/08/2002& 15$\times$400   & 515 & 1.391$-$2.085&0.6$-$0.9& $-$0.012,$+$0.017&on transit\\ 
4&15/08/2002& 15$\times$400   & 475 & 1.388$-$1.704&0.7$-$1.7& $+$0.826,$+$0.857&off transit\\
5&13/09/2002& 15$\times$400   & 485 & 1.958$-$1.385&0.8$-$1.8& $+$0.003,$+$0.026&on transit\\ 
6&20/09/2002& 15$\times$400   & 515 & 1.695$-$1.384&0.7$-$1.0& $-$0.009,$+$0.020&on transit\\ 
7&22/09/2002& 15$\times$400   & 590 & 1.446$-$1.493&0.5$-$1.0& $+$0.569,$+$0.599&off transit\\ 
\end{tabular}
\end{table*}

Since the discovery that the extra-solar planet HD209458b transits
its host star (Charbonneau et al. 2000; Henry et al. 2000; Mazeh et al. 2000),
many attempts have been made to detect its atmosphere using transmission
spectroscopy. During a transit, light that passes from the star through the
outer parts of the planet's atmosphere has impressed on it a spectrographic 
signature of the atmospheric constituents, which can be observed as an extra
absorption on top of the stellar spectrum 
(Seager \& Sasselov 2000; Brown 2001; Hubbard et al. 2001). 
Observations with the Hubble Space Telescope (HST)
using this technique have been a great success. First the Sodium D feature
was discovered by Charbonneau et al. (2002) with a 
relative depth of 0.02\%, followed by the detection of
very strong absorption features at a 5$-$10\% level in the ultra-violet 
from Hydrogen, Oxygen and Carbon (Vidal-Madjar et al. 2003; 2004). 
The latter most likely come from the `exosphere' of HD209458b, an extended 
cometary tail of gas evaporating from the planet caused by its migration  
close to the star (eg. Schneider et al. 1998).

In strong contrast, attempts to detect the atmosphere of HD209458b 
from the ground have so far been unsuccessful. So far, only upper limits
of typically 1-2\% have been reached, for the Sodium D 
feature in the optical (Brown et al. 2000; Bundy \& Marcy 2000; Moutou et al. 
2001), and  for He, CO, H$_2$O and CH$_4$ in the near-infrared (Brown et al. 2002; 
Harrington et al. 2002; Moutou et al. 2003).

Ground-based observations suffer greatly from the fact that transmission
spectroscopy relies on the comparison of spectra on and off 
transit, which are necessarily taken on different nights.
This limits the accuracy due to varying weather conditions and
instability of the instruments.
In this paper a new method for probing the atmospheres
of transiting exoplanets is presented, making use of the Rossiter effect.
Due to the rotation of the host star, a transiting planet will first
block off light from the approaching and then from the receding 
parts of the stellar surface. This results in an artificial
wobble in radial velocity, an effect first observed by Rossiter (1924) for
the eclipsing binary $\beta$ Lyrae.
The amplitude of this signal is directly proportional to the effective 
size of the transiting planet, and the wavelength dependence of this effect
can reveal atmospheric absorption features, in the same way as with
transmission spectroscopy. The advantage of this method is that 
it does not rely on accurate photometric comparison of spectra in and
out of transit, but instead depends on relative velocity shifts of stellar 
absorption lines in the same in-transit spectra.

\section{Observations, data reduction, and analysis}

\begin{figure}
\psfig{figure=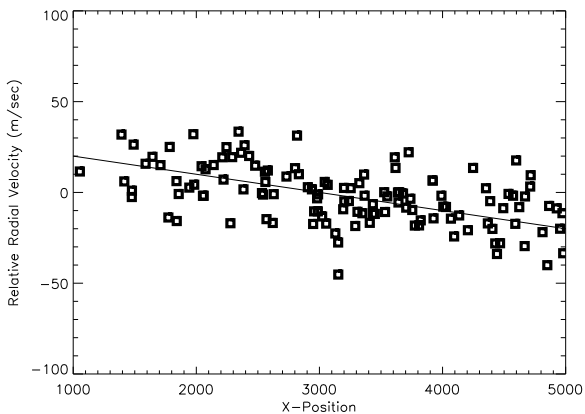,width=8cm}
\caption{\label{flexure} An example of the change in wavelength solution 
between two exposures. The x-position is the pixel position of a line 
within its order. The relative radial velocity indicates the change 
in velocity of a line between the two exposures, corrected for a
global velocity shift. This effect is removed from the data by
fitting the data points with a 3rd order Chebyshev function (solid line).}
\end{figure}

We applied this new method to an archival data set of HD209458 taken with the 
UV-Visual Echelle Spectrograph (UVES) on the Very Large Telescope.
This data was originally obtained to perform conventional atmospheric 
transmission spectroscopy (PI: Moutou et al., project 69.C-0263).
Data was collected during 7 nights (4 on, and 3 off transit) 
through UVES' red arm only, using the CD3 cross disperser with a blaze 
wavelength of 5600 \AA$ $, and image slicer nr. 3, consisting of 5 slices of 
0.3'' width each. Each night a series of 15 data frames were taken with
exposure times of 400 sec. A calibrator was observed after the 5th 
and 10th frame each time, leaving gaps of $\sim$25 min.
The observation log is given in table \ref{log}. 

The detector in the red arm of UVES consists of a mosaic of 
two 4k$\times$2k CCDs. For our analysis we concentrated on 
one CCD, covering a wavelength range of 5850 to 6550 \AA$ $ over 
16 echelle orders, and which contains the two Sodium D lines 
(5890.0 \& 5895.9 \AA) in its first order.
The data reduction was performed in a standard manner, using 
ESO's UVES data reduction pipeline, which is part of the software package 
MIDAS. 
Data from each echelle order was kept separately, with the 
output data of each frame consisting of a 2 dimensional array with
wavelength position along the x-axis (pixel size = 0.0174 \AA), 
and echelle order along the y-axis. The resolving power was
typically $\sim$110,000.
The orbital phase of HD209458b was determined for each observation using the 
orbital parameters as derived by Brown et al. (2001), and were corrected
for variations in light traveling time through the solar system.
A global phase-shift of 0.008 (40 min.) had to be applied to all data sets
to centre the Rossiter effect at zero phase. This shift corresponds to 
an error in the period derived by Brown et al. of 8 sec.

The subsequent data analysis on the 2 dimensional data 
frames was performed in IDL. 
First the spectrum in each order was fitted with a 7th order Chebyshev 
function away from absorption features, and normalised. 
Further analysis consisted of the four steps described
below. Since some of their input and/or output are interdependent they are 
performed several times before the final results were produced. 

\paragraph*{Step 1: Absolute wavelength calibration.}

The absolute wavelength calibration was determined from the 
wavelength positions of the strong H$_2$O telluric 
lines between 6275\AA$ $ and 6320\AA$ $ (in the 4th echelle order). 
First a model of the stellar 
absorption spectrum of HD209458 was produced from the velocity 
shifted and median filtered spectra obtained during the 7 nights.
This model was then appropriately shifted and subtracted from
each frame, leaving only the telluric line absorption.
The dozen strongest lines were fitted with Gaussians and their 
positions compared with 
the spectral atlas of telluric line absorption of 
Pierce \& Breckenridge (1974). 
After several iterations, necessary to obtain the 
best stellar model, the absolute wavelength calibration is found to 
be accurate to $\sim$5-10 m/sec.

\paragraph*{Step 2: Fitting of the stellar absorption lines.}

The wavelength positions of the $\sim$300 strongest absorption lines
were determined by least-square fitting of an area of 40 pixels 
centred around the minimum of each absorption line.
The accuracy of this method was estimated by measuring the standard
deviation around the mean position of the line during each night,
corrected for earth rotation and orbital motion and
changes in the wavelength solution.
The two Sodium D lines were fitted in a similar way, but 
with a 4 parameter Moffat function instead of a Gaussian.

\paragraph*{Step 3: Relative wavelength calibration.}

It was found that the relative positions of the strong emission lines 
varied much more than expected, and that these variations were a 
function of the position of a line within its order
(a typical example is shown in Figure \ref{flexure}).
A similar but generally smaller trend was also found as function of 
the echelle order. These effects are most likely caused by small 
changes in the instrument settings and/or temperature of the 
spectrograph from exposure to exposure. 
We simply corrected for these trends by fitting relative positions
of the $\sim$100 strongest lines as function of their pixel position
with a 3rd order Chebyshev function, and with a 2nd order polynomial
as function of the echelle order.
The residual scatter around the best fits are found to be 
typically 10$-$15 m/sec.
This correction in the wavelength solution is applied to all lines.

\paragraph*{Step 4: Removal of telluric line emission.}

Great care has to be taken in the removal of telluric line absorption,
especially around the Sodium D line at 5889.9 \AA.
A small change in the strength of the telluric absorption can 
introduce a significant shift in radial velocity, if it is located
within the fitting region of the stellar absorption line.
Using the best wavelength solution available, a model of the stellar 
spectrum of HD209458, as obtained above, is subtracted from the 
data, leaving only the telluric lines. Using the line list 
of Pierce \& Breckenridge (1974), the telluric absorption lines
are fitted with Gaussians, with the amplitude as only free parameter,
and are subsequently removed.

\begin{figure}
\psfig{figure=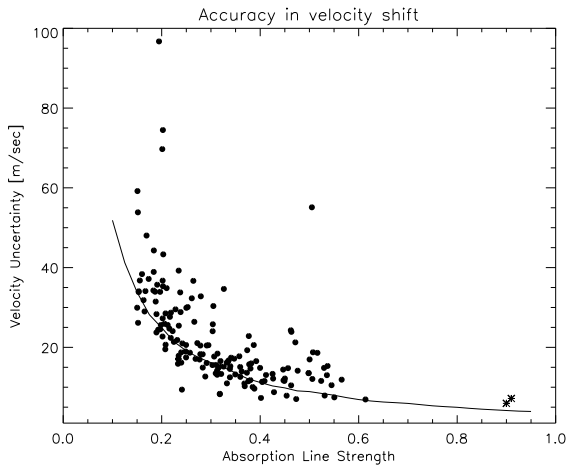,width=8cm}
\caption{\label{accuracy} 
The accuracy in radial velocity as function of the absorption line
strength for one of the data sets. This is estimated from the 
scatter around the mean position of each line.
The two Sodium D lines are indicated
by stars. The solid line indicates the expected uncertainty due to 
Poisson noise for a spectrum with a signal-to-noise ratio of 500
per spectral element. An accuracy of 5$-$10 m/sec is reached for 
the strongest lines.}
\end{figure}

The accuracy in radial velocity achieved for one of the data sets 
is shown as function of the absorption line depth in Figure \ref{accuracy}.
It shows that for the strongest lines, such as the Sodium D lines,
an accuracy of 5-10 m/sec has been reached. Monte-Carlo simulation were
performed to determine the theoretical uncertainty as function of 
absorption line strength, assuming a signal to noise ratio of 500 per 
spectral element. This is indicated by the solid line.

\section{Results}
\begin{figure}
\psfig{figure=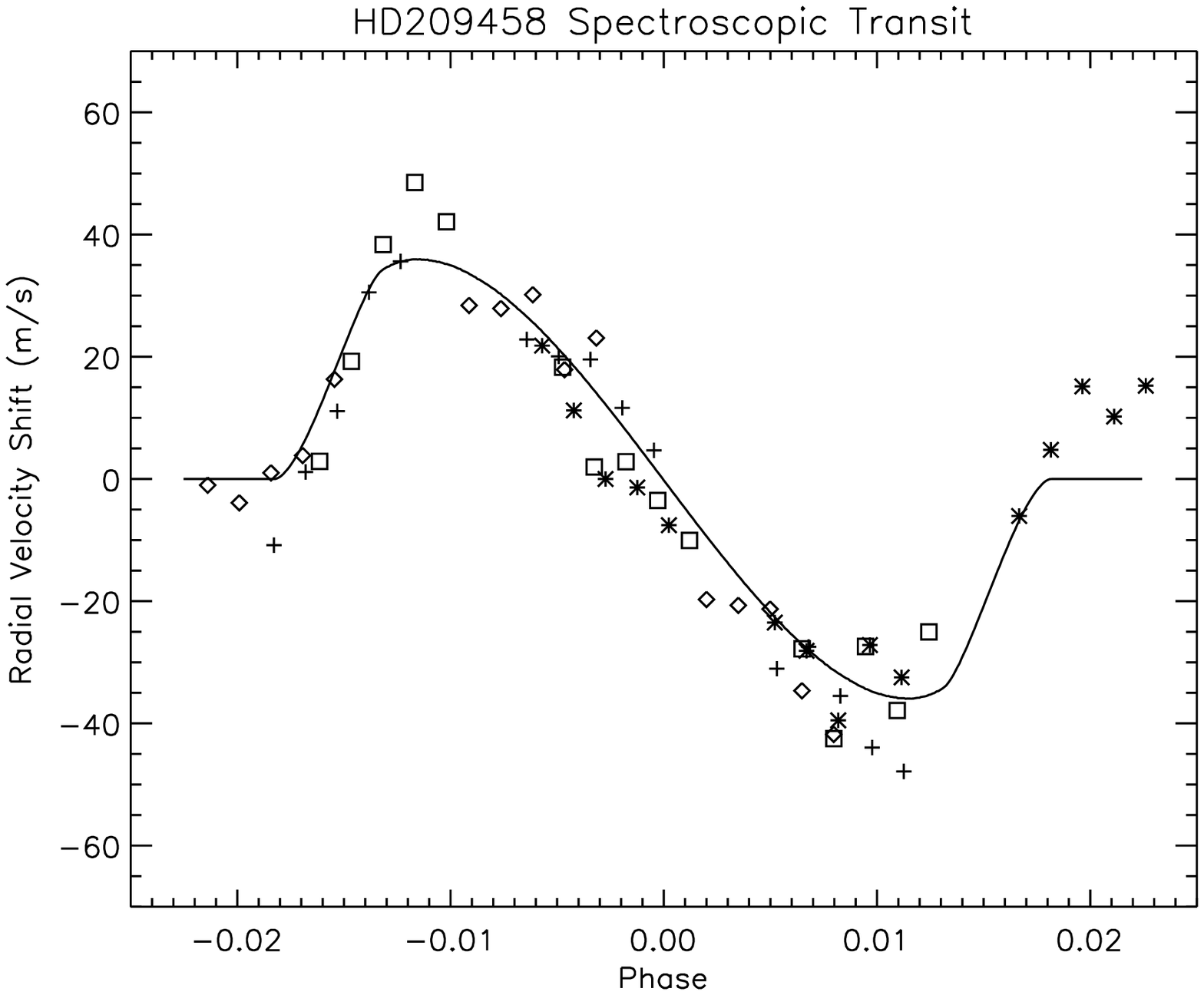,width=8cm}
\psfig{figure=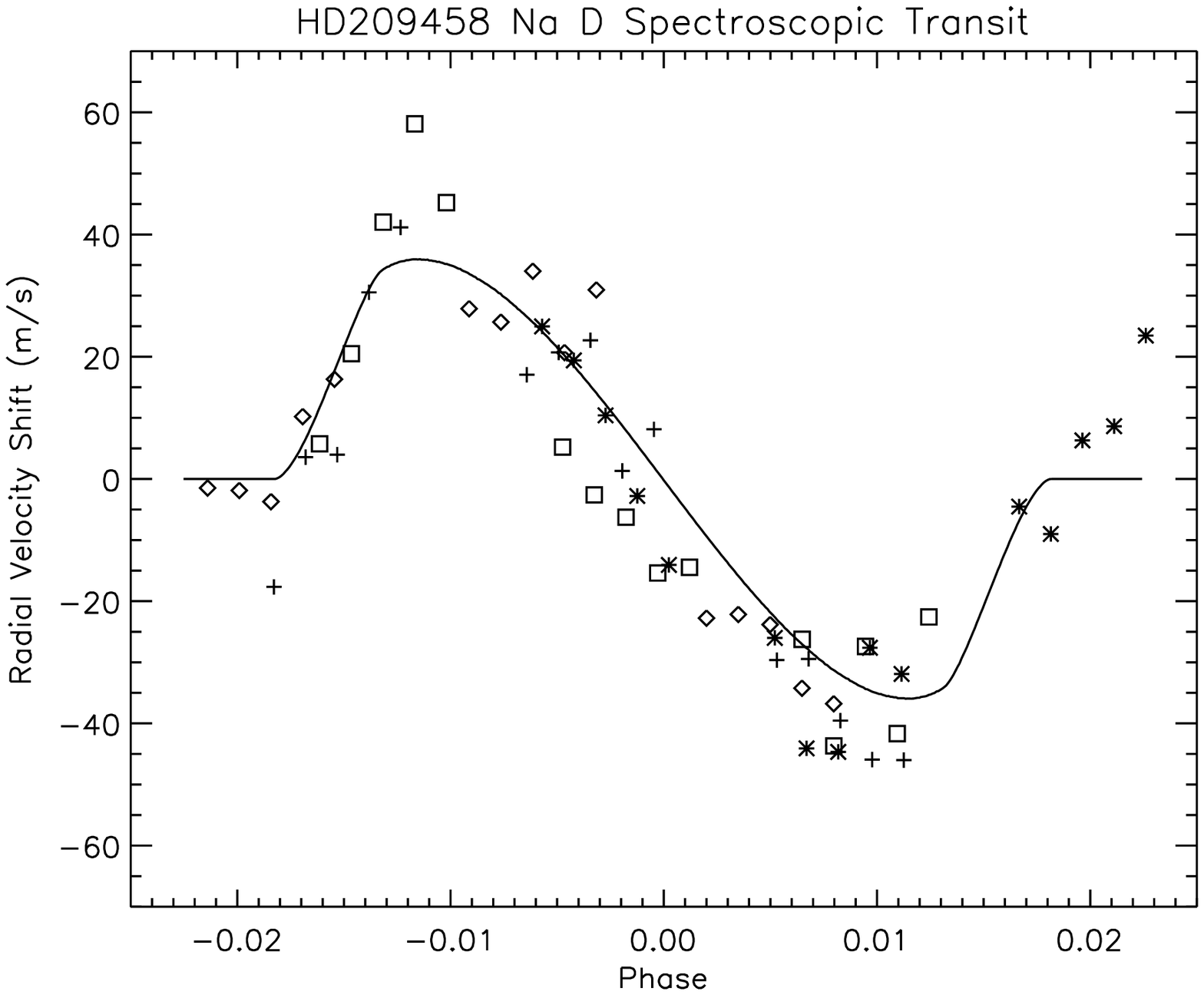,width=8cm}
\caption{\label{amplitude} The Rossiter effect as observed for HD209458,
using the weighted average of the $\sim$100 strongest
absorption lines (upper panel), and the average of the 
two Sodium D lines (lower panel).
Data from the four on-transit data sets are indicated with
boxes - set 1, diamonds - set 2, stars - set 4, and crosses - set 5.
Our best model is indicates by the solid line.}
\end{figure}

First the overall strength of the Rossiter effect of HD209458 was determined. 
The weighted mean of the radial velocities of all 
absorption lines was determined for each exposure, and subsequently
corrected for the star's motion induced by the exoplanet 
(assuming an amplitude of 85.9 m/sec; Mazeh et al. 2000), 
the motion and rotation of the Earth, and for the shifts in 
the absolute wavelength calibration as determined above. 
The same corrections were applied
to the radial velocities of the two sodium D lines.
The 4 on-transit data sets were then merged by applying small 
global $0-5$ m/sec 
offsets to the data to minimise the scatter in the combined velocity 
curve. The merged velocity curves are shown in figure \ref{amplitude},
which are fitted to models of the Rossiter effect. Our models are 
constructed in a similar way as in Queloz et al. (2000),
by simulating the transit of a dark object over a limb-darkened star.
A spherical star with uniform rotation is considered, with the orbital 
motion of the planet and the stellar rotation set in the same direction.
The angle between the orbital plane and the equatorial plane is 
assumed to be zero, and the impact parameter is set to 0.5, corresponding
to the orbital inclination of 86.6$^\circ$ as determined by 
Brown et al. (2001). The star is divided into 200,000 cells, with the spectrum
of each cell modeled by a Gaussian absorption line with $\sigma = 4.0$ km/sec.
The overall integrated spectrum is made by summation of the cells free of 
the planet along the line of sight.
 
The aim of this modeling is to set the amplitude scale
of the Rossiter effect relative to the depth of the photometric transit.
Limb darkening is assumed to 
be of the form $B_\mu = 1 -\epsilon(1-\mu)$, where $\mu$ is the cosine 
of the angle between the line of sight and the normal to the local stellar 
surface, and $\epsilon$ is assumed to be 0.58 (Deeg et al. 2001). 
The influence of wavelength dependent limb-darkening is discussed below.
Assuming a ratio of planet to star radius of 0.118 (Brown et al. 2001),
the overall velocity curve is best fitted with a stellar $v\ sin\ i$ of 
4.5 km/sec, resulting in an amplitude of the Rossiter effect of about 
38 m/sec (see figure \ref{amplitude}). This amplitude corresponds to 
the photometric transit depth of 1.6\% (Brown et al. 2001),
meaning that an excess of 1 m/sec compares to a signal for conventional 
atmospheric transmission spectroscopy of 0.04\%.

\subsection*{The excess of the Rossiter effect in the Sodium D lines}

\begin{figure*}
\hbox{
\psfig{figure=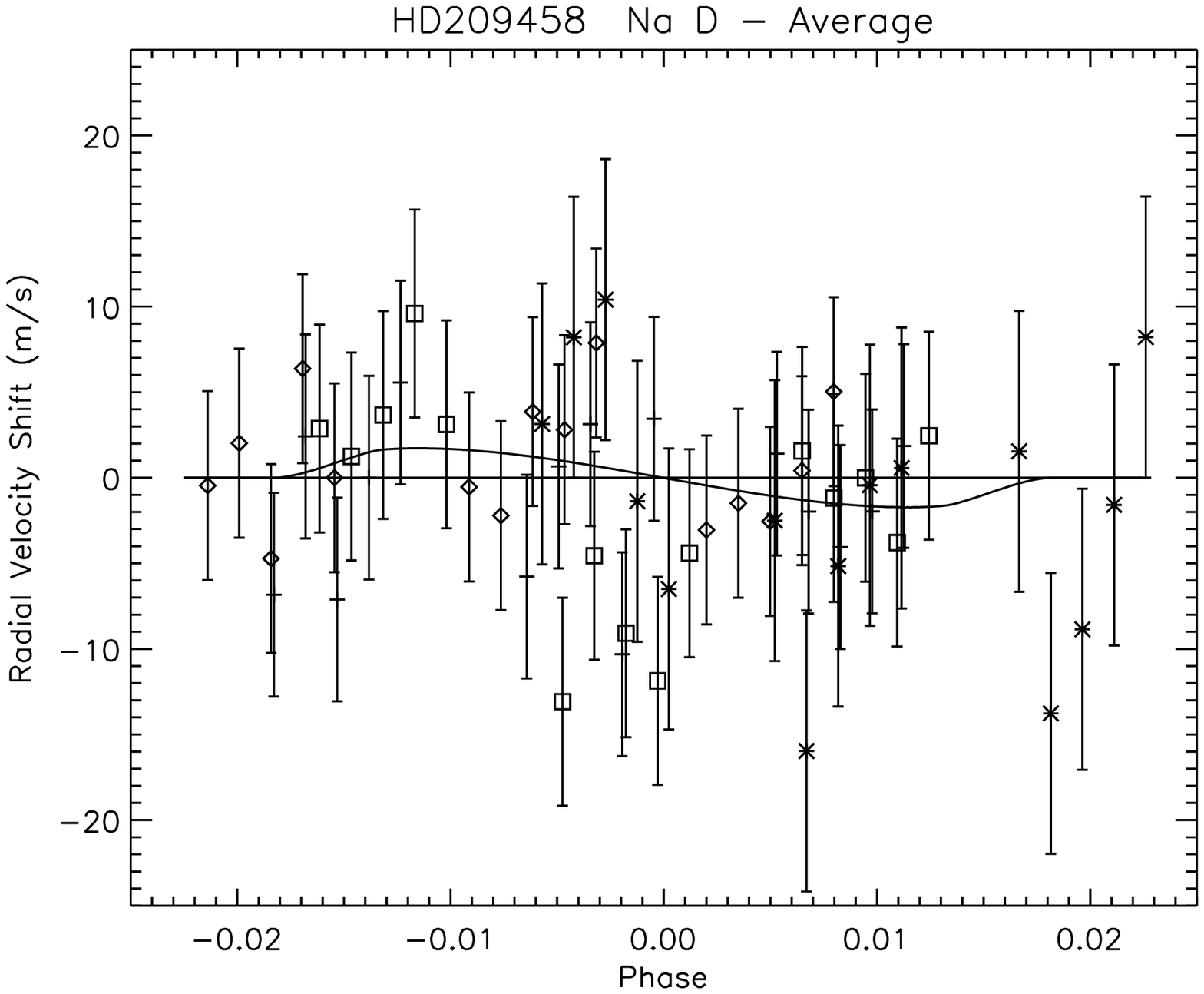,width=8cm}
\psfig{figure=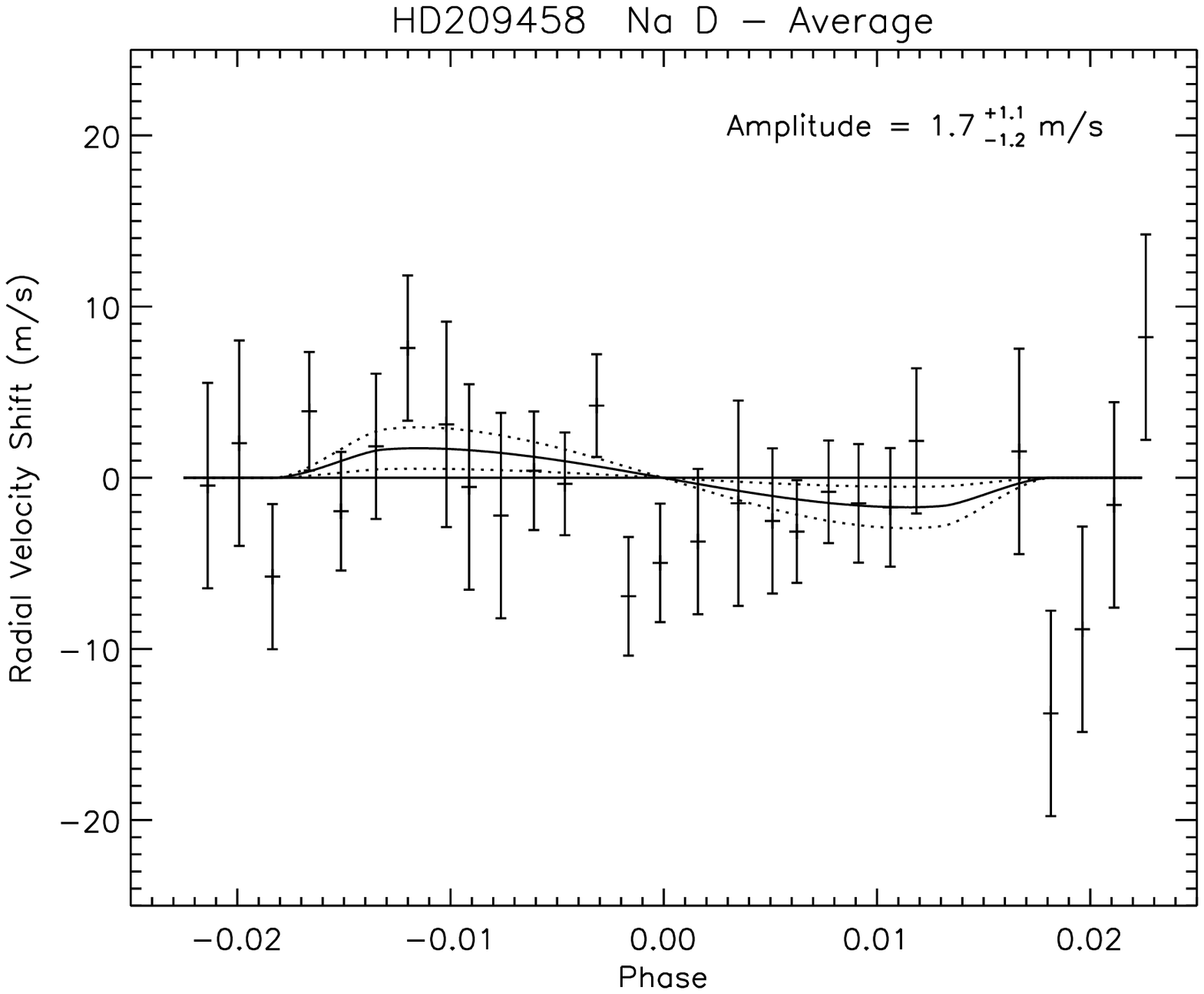,width=8cm}
}
\caption{ \label{relative} 
The excess of the Rossiter effect in Sodium D, using the individual
data points (left panel) and binned data (right panel).
The best fit gives an amplitude of 1.7$^{+1.1}_{-1.2}$ m/sec,
and is shown by the solid line. The 1$\sigma$ confidence limits are 
indicated by the dotted lines.
This corresponds 
to a photometric signal using conventional transmission spectroscopy of 
7$\pm$5$\times10^{-4}$.
The symbols are as in figure \ref{amplitude}.}
\end{figure*}

The mean radial velocity of each exposure were then subtracted from 
that of the Sodium D lines to reveal any possible excess in the 
Rossiter effect. 
The global velocity shifts applied above cancel out,
but  new global velocity shifts were applied to each data set, to set the mean 
relative velocity shift of all 15 exposures to zero. Ideally, the observations 
should have cover a substantial period before and after the transit to 
set the zero-point independently from the in-transit data (see below).
The velocity shifts of the Sodium D lines compared to the average of 
all other lines is shown in figure \ref{relative}. 
The scatter in the four data sets is 6.5, 3.7, 7.6, and 4.7 m/sec 
respectively. The right panel shows the data binned together,
using time-bins of 450 seconds. 
The dispersion around zero in this binned
data set is 4.7 m/sec.
The data set is fitted with the model described above, with all parameters
fixed except the amplitude of the signal. The best fit gives
an amplitude of $1.7^{+1.1}_{-1.2}$ m/sec, corresponding to a 
photometric signal using conventional atmospheric transmission spectroscopy
of 7$\pm$5$\times10^{-4}$.

\section{Discussion and conclusions}

The analysis of the UVES archive data shows that the Rossiter effect
provides a powerful tool to probe the atmospheres of transiting exoplanets.
Although no significant signal has been detected for Sodium D in HD209458b, 
it is the first time that ground-based observations result in an 
accuracy of $<$0.001. Previous ground-based observations have 
reached accuracies of only 1$-$2\%, corresponding to an excess in the 
Rossiter effect of 20$-$50 m/sec, clearly inferior to the precision
achieved here.

The HST Transmission spectrum of HD209458b obtained by 
Charbonneau et al. (2002) revealed an excess in absorption 
in Sodium D of 0.00023 compared to the surrounding continuum, 
implying a relative increase of $\sim$1.8\%.
This indicates that for this particular absorption feature an excess in the 
Rossiter effect of $\sim$0.7 m/sec can be expected, 
making its detection very challenging for the current instrumentation. 
However, it is important to realise that 
the excess in absorption as determined with the HST was measured over
a bandwidth of 12\AA. In contrast, line-to-line variations in 
the Rossiter effect will be most sensitive to planetary absorption on 
scales of the half-power width of the stellar lines (a fraction of an \AA ngstrom). Most models of transmission spectra (eg. Brown 2001) indicate that 
on this scale the absorption signal for Sodium D could be 2$-$5 times higher 
(1$-$3 m/sec).

The archival data sets used in this paper are not optimal for measuring the
Rossiter effect. First of all, since a significant fraction
of the time was devoted to calibrators, typically only 60\% of the on-transit
time was spent on target, and on several occasions the periods that the 
amplitude of the Rossiter effect is expected to be the strongest were missed. 
In addition, the current analysis is significantly hampered by the fact
that no or hardly any time was spent on target directly before or after 
each transit.
This was often dictated by the timing of the transit and visibility of 
HD209458 from Paranal. For the most optimal transits, HD209458 can be 
observed from 1 hour before to 1 hour after the overall transit with the VLT.
In those cases the zero-point in velocity can be determined independently from
the on-transit data, and one could also correct for any possible residual  
drifts in radial velocity over the transit, eg. due to residual 
telluric absorption line features or inaccuracies in flat fielding.
Overall, this should result in an increase in precision of a factor
$2-3$. It means that, if the current value is believed, 
a 3$-$5 $\sigma$ detection of Sodium D should be possible over a couple of 
transits. 

\subsection*{Wavelength variations in limb-darkening}

The exact shape and amplitude of the Rossiter effect are dependent 
on the stellar limb darkening. So far we have not taken into account that 
limb darkening decreases as function of wavelength. 
Furthermore, the limb darkening effect also changes 
differently over the profile of each absorption line. 
Here we assess the possible influence of both effects.
The wavelength dependence of the limb darkening of HD209458 has been
observed by Deeg et al. (2001), with $\epsilon$
changing from 0.62 at 5900 \AA$ $ to 0.58 at 6180 \AA.
To estimate the influence of this effect we assumed a linear wavelength 
dependence of $\epsilon(\lambda) = 1.46-\lambda/7000\AA$,
and calculated the change in radial velocity for each line relative to that 
of Sodium using the model described above. 
By incorporating this wavelength dependent limb-darkening, the
excess of the Rossiter effect in the Sodium D lines is increased by 
$\sim$0.2 m/sec. 

The variations in limb darkening across solar absorption lines is 
studied in great detail by Pierce \& Slaughter (1982) for Sodium D, and 
by Balthasar (1988) for a whole range of bright absorption lines, by
measuring changes in the absorption line depths across the solar disk.
The strength of this effect is dependent on the mechanism of formation of 
the line. Most absorption lines weaken towards the limb of the 
Sun, meaning that the limb darkening in the centre of the lines is less
than in the surrounding continuum. Assuming that this effect is similar 
for HD209458 as for the Sun, our modeling indicates that these variation 
contribute at a level of again $\sim$0.2 cm/sec. 
Although this is insignificant for these data, 
it means that future observation aimed at measuring variations in the 
Rossiter effect at much higher precision should take these effects into 
account.

\subsection*{Future applications for this method}

We believe that the new method presented here 
to probe atmospheres of transiting exoplanets 
using the Rossiter effect, has great potential. 
Specifically targeted observations, rather than the archival 
data used here, can increase the accuracy further by a factor 2$-$3.
Furthermore, in the near-infrared, where the atmospheric
absorption features due to H$_2$O, CO, and CH$_4$ cover many 
stellar lines, a great improvement over current observations can 
be reached. Although confusion with water-vapor and
methane in the earth atmosphere will complicate the analysis.
In addition, anticipating the discovery of exoplanets transiting
bright stars of either later spectral type (with many more stellar lines
to average), or with faster stellar rotations (resulting in a Rossiter
effect with a higher amplitude), a further significant increase in 
accuracy can be expected.

\section*{Acknowledgements}
I would like to thank Philip Best for  useful discussions and for 
carefully reading the manuscript.
This research is based on observations made with the European Southern 
Observatory
telescopes obtained from the ESO/ST-ECF Science Archive Facility.


\begin{thebibliography}{99}
\bibitem{} Balthasar H., 1988, A\&AS 72, 473
\bibitem{} Brown T.M., Butler R.P., Charbonneau D., Noyes R.W., Sasselov D., 
            Libbrecht K.G., Marcy G.W., Seager S., Vogt S.S., 2000, 
            197th AAS Meeting, 11.05; BAAS, Vol. 32, p.1417
\bibitem{} Brown T.M., 2001, ApJ 553, 1006 
\bibitem{} Brown T.M., Charbonneau D., Gilligand R.L., Noyes R.W., Burrows A.,
            2001, ApJ 552, 699
\bibitem{} Brown T.M., Libbrecht K.G., Charbonneau D., 2002, PASP 114, 826
\bibitem{} Bundy K.A., Marcy G.W., 2000, PASP 112, 1421
\bibitem{} Charbonneau D., Brown T.M., Latham D.W., Mayor M., 2000, ApJ 529, 45
\bibitem{} Charbonneau D., Brown T.M., Noyes R.W., Gilligand R.L., 2002,
            ApJ 568, 377
\bibitem{} Deeg H.J., Garrido R., Claret A., 2001, NewA, 6, 51
\bibitem{} Harrington J., Deming D., Matthews K., Richardson L.J., Rojo P.,
             Steyert D., Wiedemann G., Zeehandelaar D., 2002,
             201st AAS Meeting, 46.04; BAAS, Vol. 34
\bibitem{} Henry G.W., Marcy G.W., Butler R.P., Vogt S.S., 2000, ApJ 529, L41
\bibitem{} Hubbard W.B., Fortney J.J., Lunine J.I., Burrows A., Sudarsky D.,
            Pinto P., 2001, ApJ 560, 413
\bibitem{} Mazeh T., Naef D., Torres G., Latham D.W., Mayor M., Beuzit J.-L.,
            Brown T., Buchlave L., Burnet M., Carney B.W. et al. 2000, ApJ 
            532, L55
\bibitem{} Moutou C., Coustenis A., Schneider J., St Gilles R., Mayor M., 
            Queloz D., Kaufer A., 2001, A\&A 371, 260 
\bibitem{} Moutou C., Coustenis A., Schneider J., Queloz D., Mayor M.
             2003, A\&A 405, 341
\bibitem{} Pierce A.K., Breckenridge J.B., Kitt Peak National Observatory 
            Contribution, Tucson: Kitt Peak National Observatory, 1974
\bibitem{} Pierce A.K., Slaughter C., 1982, ApJS 48, 73
\bibitem{} Queloz D., Eggenberger A., Mayor M., Perrier C., Beuzit J.L,
            Naef D., Sivan J.P., Udry S., 2000, A\&A 359, L17
\bibitem{} Rossiter R.A., 1924, AJ 60, 15
\bibitem{} Seager S., Sasselov D.D., 2000, ApJ 537, 916 
\bibitem{} Schneider J., Rauer H., Lasota J.P., Bonazzola S., Chassefiere E.,
            1998, in ``Brown dwarfs and extrasolar planets'', 
            eds R. Rebolo, E.L. Martin; M.R. Zapatero Osorio, 
            ASP Conference Series 134, 241
\bibitem{} Vidal-Madjar A, Lecavelier des Etangs A., D\'{e}sert J.-M., 
            Ballester G.E., Ferlet R., H\'{e}brard G., Mayor M., Nature 422,
            143
\bibitem{} Vidal-Madjar A, D\'{e}sert J.-M., Lecavelier des Etangs A.,
            H\'{e}brard G., Ballester G.E., Ehrenreich D., Ferlet R.,
            McConnell J.C., Mayor M., Parkinson C.D., 2004, ApJL submitted

\bibitem{} 
\bibitem{} 

\end{thebibliography}
\end{document}